\documentclass[fleqn,usenatbib]{mnras}

\usepackage{newtxtext,newtxmath}
\usepackage[T1]{fontenc}

\DeclareRobustCommand{\VAN}[3]{#2}
\let\VANthebibliography\thebibliography
\def\thebibliography{\DeclareRobustCommand{\VAN}[3]{##3}\VANthebibliography}

\usepackage{graphicx}	% Including figure files
\usepackage{amsmath}	% Advanced maths commands
\usepackage{color}

\title[Anti-correlation of Swift J1727.8--1613]{Energy-dependent Optical/Near-infrared and X-ray Correlations in Swift J1727.8--1613}
\author[Ruican Ma et al.]{
Ruican Ma, $^{1}$\thanks{E-mail: R.Ma@soton.ac.uk (UoS)}
Federico Vincentelli, $^{2}$ 
Diego Altamirano, $^{1}$
Alexandra Veledina, $^{3,4}$
Poshak Gandhi, $^{1}$ 
\newauthor
Piergiorgio Casella, $^{5}$
Tariq Shahbaz, $^{6,7}$
Sian Woahene-Demehin $^{1}$
\\
$^{1}$School of Physics and Astronomy, University of Southampton,
Highfield, Southampton, SO17 1BJ, UK\\
$^{2}$Centre for Fluid and Complex Systems, Coventry University, CV1 5FB, Coventry, UK\\
$^{3}$Department of Physics and Astronomy, FI-20014 University of Turku, Finland \\
$^{4}$Nordita, KTH Royal Institute of Technology and Stockholm University, Hannes Alfv\'ens v\"ag 12, SE-10691 Stockholm, Sweden \\
$^{5}$INAF-Osservatorio Astronomico di Roma, Via Frascati 33, I-00076 Monte Porzio Catone (RM), Italy \\
$^{6}$Instituto de Astrofisica de Canarias (IAC), Calle Via Lactea s/n E-38200 La Laguna, Tenerife, Spain \\
$^{7}$Departamento de Astrofisica, Universidad de La Laguna (ULL), Av. Astrofisico Francisco Sanchezs/n, E-38206 La Laguna, Tenerife, Spain \\
}

\date{Accepted XXX. Received YYY; in original form ZZZ}

\pubyear{\the\year{}}

\begin{document}
\label{firstpage}
\pagerange{\pageref{firstpage}--\pageref{lastpage}}
\maketitle

\begin{abstract}
We present a timing analysis of the black hole transient Swift J1727.8--1613 during its intermediate state. 
We use coordinated broadband X-ray observations from \textit{Insight}-HXMT (2--150 keV), together with optical data from ULTRACAM ($g_s$ and $i_s$ bands) and near-infrared data from HAWK-I ($K_s$ band), obtained on 2023 September 9. As shown by previous studies, the Fourier power spectrum shows a strong quasi-periodic oscillation (QPO) in the $K_s$, $i_s$ and X-ray bands.
Cross-correlation analysis reveals a complex coupling between the optical/near-infrared (OIR) and X-ray emission, including a delayed optical anti-correlation, a strong infrared correlation, and a pronounced dependence of these features on X-ray energy, suggesting multiple Comptonisation regions.
In contrast, the lag properties do not change at the QPO frequency, displaying an OIR lag of $\sim$60–80 ms up to 150 keV. 
We discuss these results in the context of small-scale jet and hot accretion flow scenarios.
\end{abstract}


%%%%%%%%%%%%%%%%%%%%%%%%%%%%%%%%%%%%%%%%%%%%%%%%%%

%%%%%%%%%%%%%%%%% BODY OF PAPER %%%%%%%%%%%%%%%%%%

\section{Introduction}

Black hole transients (BHTs) constitute a subclass of low-mass black hole X-ray binaries (BHXBs) that remain quiescent for extended periods before undergoing luminous outbursts with multiwavelength emission spanning X-ray, optical, infrared (IR) and radio bands \citep[e.g.,][]{Done2007, Hynes2003, Fender2001}. 
In black hole X-ray binaries, lower-energy X-rays are generally more closely linked to the accretion disk, whereas higher-energy X-rays are mainly shaped by Comptonisation and possible jet-related emission \citep{Done2007, Remillard2006}.
The optical and near-infrared (OIR) emission observed in BHTs is generally attributed to a combination of X-ray reprocessing, synchrotron radiation from the hot flow, and/or jet (e.g., \citealt[][]{Hynes2003, Veledina2011, Malzac2014, Poutanen2014, Uttley2014}, and recent studies \citealt[][]{You2023, Du2025, Fan2026}).

Multiwavelength timing analysis is a powerful tool for probing the origin of fast variability in BHXBs \citep[e.g., see the review of ][]{Uttley2014}. Positive OIR lags of a few hundred milliseconds with respect to X-ray have been observed in several BHXBs through cross-correlation function (CCF) analyses, such as GX 339--4 \citep{Gandhi2008, Casella2010}, V404 Cyg \citep{Gandhi2017} and MAXI J1820+070 \citep{Paice2019, Paice2021}. 
On the short time scale, the sub-second OIR positive lag with a correlation can be interpreted within the jet internal-shock model (ISHEM), where variable-speed shells are launched along the jet \citep{Malzac2018}. Faster shells catch up with slower ones, producing shocks that accelerate electrons and generate OIR synchrotron emission; the lag then traces the propagation time from the inner accretion flow to the OIR-emitting region \citep[e.g.,][]{Casella2010, Malzac2014, Malzac2018}. Further frequency-resolved cross-spectral studies have confirmed the few hundred milliseconds OIR positive lag at a relatively high frequency (e.g., > 0.2\,Hz), which is typically accompanied by relatively high coherence \citep{Gandhi2010, Vincentelli2019, Paice2019}. 
On longer timescales (of the order of seconds to tens of seconds), the ISHEM model may also provide a natural explanation for anti-correlations between the X-ray and OIR emission. In this picture, a temporary decrease in the jet velocity can reduce the efficiency of internal shocks, delaying the synchrotron response until the velocity fluctuations propagate downstream. The OIR emission would then be more closely linked to variations in the jet velocity than to the instantaneous X-ray flux, potentially producing a long-timescale anti-correlation accompanied by a short-timescale OIR positive lag \citep{Malzac2018}.
On even longer timescales, an optical lag of $\sim$17 days relative to the hard X-rays was reported in MAXI J1820+070 and attributed to the thermal-viscous response of the outer disk rather than the jet \citep{You2023}, indicating that lags on different timescales trace distinct physical processes.

X-ray/OIR CCFs have also shown different kinds of features: early observations of XTE J1118+480 \citep{Kanbach2001,Malzac2004} and later of Swift J1753.5--0127 \citep{Hynes2009,Veledina2017} revealed a strong anti-correlation at negative optical lags, followed by a positive optical response. This behaviour, often referred to as the ``precognition dip'', has been interpreted in the context of the hot-flow model. 
In this framework, non-thermal OIR emission is expected to arise from synchrotron radiation produced by hybrid electron populations in the hot flow \citep{PoutanenVeledina2014}. This component can be anti-correlated with the X-ray Comptonized emission as the source luminosity increases, while disk reprocessing may also produce a delayed positive optical response on timescales of a few seconds \citep{Veledina2011,Veledina2013a,Veledina2017,Veledina2018}.
More recent observations have revealed a more complex picture, including nearly symmetric optical/X-ray anti-correlations in MAXI J1820+070 \citep{Paice2019} and asymmetric optical delays in MAXI J1535--571 \citep{Vincentelli2021}. These results suggest that OIR/X-ray timing behaviour in BHTs is diverse and may reflect the interplay of multiple emission or variability processes.

In addition, X-ray quasi-periodic oscillations (QPOs) are commonly observed in BHTs and provide a powerful probe of the innermost accretion flow close to the black hole \citep[e.g.,][]{Belloni2002, Ingram2019}. 
Their physical origin has long been debated, and a number of models have been proposed, including Lense-Thirring precession of the hot inner flow or a small-scale jet, as well as coupled disk-corona oscillations \citep[e.g.,][]{Fragile2007, Ingram2009, Ma2021, Mastichiadis2022, Ma2023}. 
QPOs have also been detected in the OIR bands in several BHTs, such as Swift J1753.5--0127 \citep{Durant2009, Veledina2015}, GX 339--4 \citep{Motch1983, Gandhi2010,Kalamkar2016,Vincentelli2019}, MAXI J1535--571 \citep{Vincentelli2021}, and MAXI J1820+070 \citep{Thomas2022}. 
The detection of QPOs across multiwavelength is particularly valuable, as it provides a direct way to test whether the X-ray and OIR variability are driven by a common physical mechanism. 
In this context, geometrical models involving Lense-Thirring precession of the hot flow or a small-scale jet provide a plausible framework for interpreting multiwavelength QPOs \citep[e.g.,][]{Veledina2013b, Thomas2022}.

Most multiwavelength timing studies of BHXBs have focused on correlations between OIR emission and X-ray observations from \textit{RXTE}, \textit{NICER}, and \textit{XMM}-Newton. However, these X-ray bands often include mixed contributions from the accretion disk, the Comptonising corona, and possibly the jet. 
Hard X-ray observations, such as those obtained with \textit{Insight}-HXMT (hereafter HXMT), are less affected by disk emission and can therefore provide a cleaner view of the innermost hot flow and possible jet-base activity \citep[e.g.,][]{You2021, Ma2021, You2026}. 
In this work, we study the BHT Swift J1727.8--1613 by combining coordinated OIR coverage with hard X-ray observations from HXMT. This provides a good opportunity to explore how OIR variability is linked to the innermost accretion flow and jet activity.

Swift J1727.8--1613 was identified as a new Galactic transient on 23 August 2023, reaching a peak flux of $\sim$7.6\,Crab in the 15--50 keV band \citep{Palmer2023}. The source has an estimated mass exceeding $3.12 \pm 0.10\,M_{\odot}$ and is located at a distance of $3.4 \pm 0.3$\,kpc \citep{Mata2025}. Observational evidence from radio and X-ray properties further suggests that the system is viewed at a moderate inclination \citep[e.g.,][]{Mata2025, Yang2024, Ma2025}. 
Broadband X-ray studies reveal strong energy dependence in QPO properties and the high-frequency hump \citep{Yu2024, Yang2024, Xu2025, Li2026}, 
while spectral analyses indicate multiple Comptonisation components and dynamic disk-corona coupling \citep{Liu2024, Ma2025, He2025, Chand2026}. 
Polarimetric results suggest a radially extended corona in the hard-intermediate state \citep{Veledina2023,Ingram2024}, although no significant QPO-phase modulation in polarization has been detected \citep{Zhao2024}. Multiwavelength timing studies report OIR QPOs up to 4.2\,Hz and an optical lag of $\sim$70\,ms relative to soft X-rays \citep{Vincentelli2025}. In this work, we extend the multiwavelength timing analysis of Swift J1727.8--1613 into the hard X-ray band, providing new constraints on the origin of high-energy emission and the geometry of the inner accretion flow.

\section{Data collection} 
\label{sec:data_red}

We analyse simultaneous X-ray, optical, and near-infrared observations of Swift J1727.8--1613 obtained on 9 September 2023. 
The X-ray data were collected by the broadband (1--250\,keV) \textit{Insight}-HXMT \citep{Zhang2020}, which consists of the LE (1--15 keV; \citealt{Chen2020}), ME (5--30 keV; \citealt{Cao2020}), and HE (20--250 keV; \citealt{Liu2020}) instruments, providing time resolutions of 1\,ms, 240\,$\mu$s, and 4\,$\mu$s, respectively. We use the observation from 2023-09-08T21:38 to 2023-09-09T02:01 (ObsID P061433800907), processed with the standard \textit{hpipeline} of the HXMT Data Analysis software ({\sc hxmtdas}) v2.06, selecting only small-FOV detector events. Further details of the reduction are given in \citet{Ma2025}.
Barycentric correction was applied using \textit{hxbary} tool in {\sc hxmtdas}, adopting the same source coordinates (R.A. = 261.930$^{\circ}$, Dec. = --16.205$^{\circ}$) and the JPL DE-430 ephemeris as used in the optical and infrared analysis of \citet{Vincentelli2025}. 

Optical data were obtained with {\sc ultracam} on the 3.58\,m New Technology Telescope (NTT) at La Silla \citep{Dhillon2007}, providing simultaneous $u_s$ ($\sim$300--400\,nm), $g_s$ ($\sim$400--550\,nm), and $i_s$ ($\sim$690--840\,nm) photometry. The observations were carried out from 2023-09-08T23:56 to 2023-09-09T03:31 in drift mode, with a time resolution of $\sim$16\,ms.
We focus on the $i_s$ and $g_s$ bands, as $u_s$ is too faint for reliable analysis. Data reduction was performed using the {\sc hipercam} pipeline, including bias subtraction, flat-fielding, and aperture photometry with seeing-scaled apertures (see \citealt{Vincentelli2025} for details).

The NIR observations were conducted in the $K_s$ band (2.2\,$\mu$m) using HAWK-I mounted on the Very Large Telescope at Cerro Paranal, Chile \citep{Pirard2004}. The observations were performed in Fast Photometry mode, providing a time resolution of 125\,ms. The data were reduced using an adapted ULTRACAM pipeline, and aperture photometry was performed with a seeing-dependent aperture; the target light curve was further normalized by a nearby reference star to correct for atmospheric and seeing variations (see \citealt{Vincentelli2025} for details). A summary of the observational information is provided in Tab.~\ref{tab:obs_info}.

\begin{figure*}
%\centering
\includegraphics[width=0.9\textwidth]{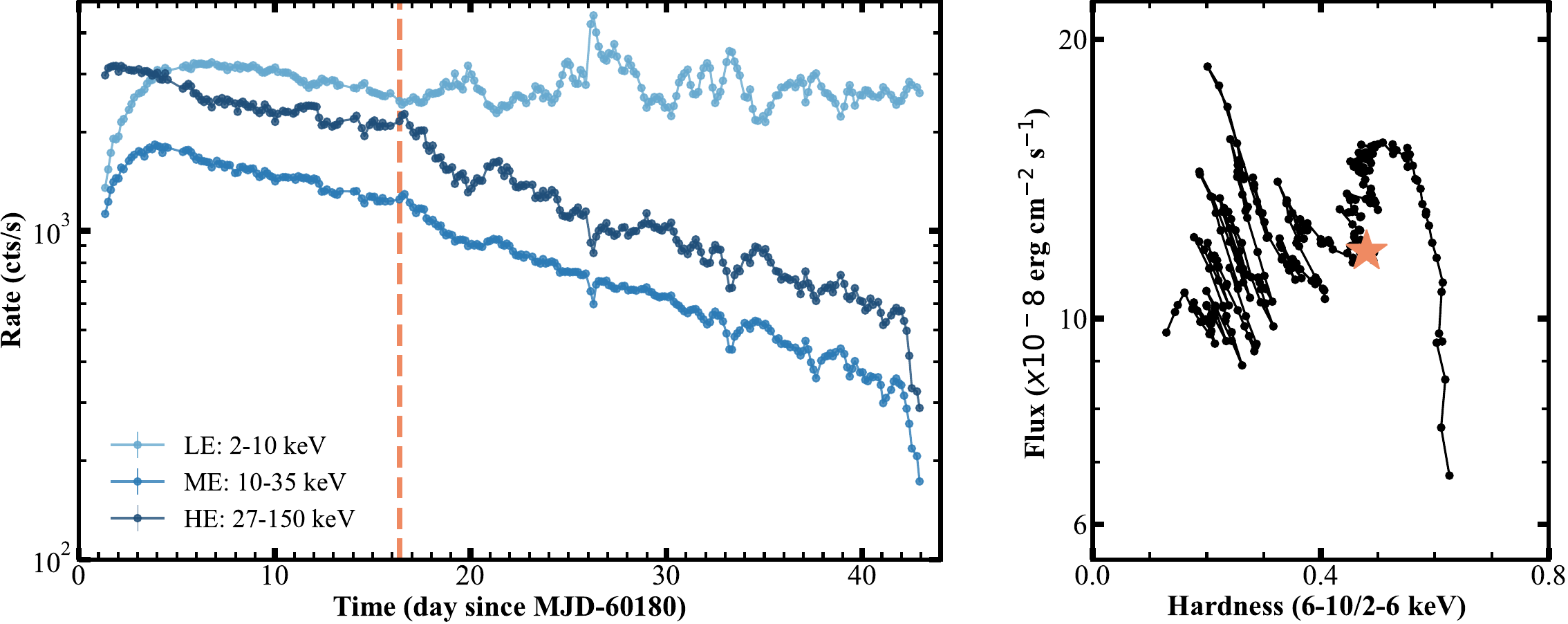} \\%column
\caption{Left panel: HXMT light curves of Swift J1727.8--1613 in the LE (2--10\,keV; light blue), ME (10--35\,keV; medium blue) and HE (27--150\,keV; dark blue) bands, respectively. The orange vertical line marks the simultaneous {\sc ultracam} observation analyzed in this work. Right panel: HID of the source. The hardness is defined as the photon count-rate ratio between the 6--10\,keV and 2--6\,keV bands, while intensity corresponds to the unabsorbed flux in the 2--10 keV. The orange star marks the observation studied here, taken during the hard-intermediate state.}
\label{fig:lc}
\end{figure*}

The left panel of Fig.~\ref{fig:lc} shows the HXMT light curves in the LE (2--10\,keV; light blue), ME (10--35\,keV; medium blue) and HE (27--150\,keV; dark blue) energy bands. The orange line indicates the time interval of the {\sc ultracam} and HAWK-I observations used in this work. The corresponding hardness intensity diagram (HID) is presented in the right panel, showing that the source was in the hard intermediate state during our observations. More details on the state classification can be found in \citet{Ma2025}.

%%%%%%%%%%%%%%%%%%%%%%%%%%%%%%%%%%%%%%%%%%%%%%%%%%%%%%%%%%%%%%%%%%%%%%%%%
\begin{table*}
\centering
\begin{tabular}{lccc}
\hline
ObsTime (UTC) & Instruments & Wavelength Band & ObsID \\ \hline
09-09T01:19 -- 09-09T01:52 & HAWK-I @VLT & $K_s$ (2.2\,$\mu$m) & 112.2615.001 \\
09-08T23:56 -- 09-09T03:31 & ULTRACAM@NTT & $i_s$ (771\,nm), $g_s$ (473.2\,nm)& NA \\
09-08T21:38 -- 09-09T02:01 & $Insight$-HXMT & LE (2--10\,keV), ME (10--35\,keV), HE (27--150\,keV) & P061433800907 \\
 \hline
\end{tabular}
\caption{Summary of the OIR and X-ray observational data for Swift J1727.8--1613. Columns, from left to right, list the observation date in 2023, instrument, wavelength band, and observation ID. The simultaneous overlap with LE/ME/HE is 1720\,s/2340\,s/2300\,s for ULTRACAM and 990\,s/1400\,s/1370\,s for HAWK-I.}
\label{tab:obs_info}
\end{table*}
%%%%%%%%%%%%%%%%%%%%%%%%%%%%%%%%%%%%%%%%%%%%%%%%%%%%%%%%%%%%%%%%%%%%%%%%%

%%%%%%%%%%%%%%%%%%%%%%%%%%%%%%%%%%%%%%%%%%%%%%%%%%%%%%%%%%%%%%%%%%%%%%%%%
%%%%%%%%%%%%%%%%%%%%%%%%%%%%%%%%%%%%%%%%%%%%%%%%%%%%%%%%%%%%%%%%%%%%%%%%%

%%%%%%%%%%%%%%%%%%%%%%%%%%%%%%%%%%%%%%%%%%%%%%%%%%%%%%%%%%%%%%%%%%%%%%%%%
\begin{figure}
\centering
\includegraphics[width=\columnwidth]{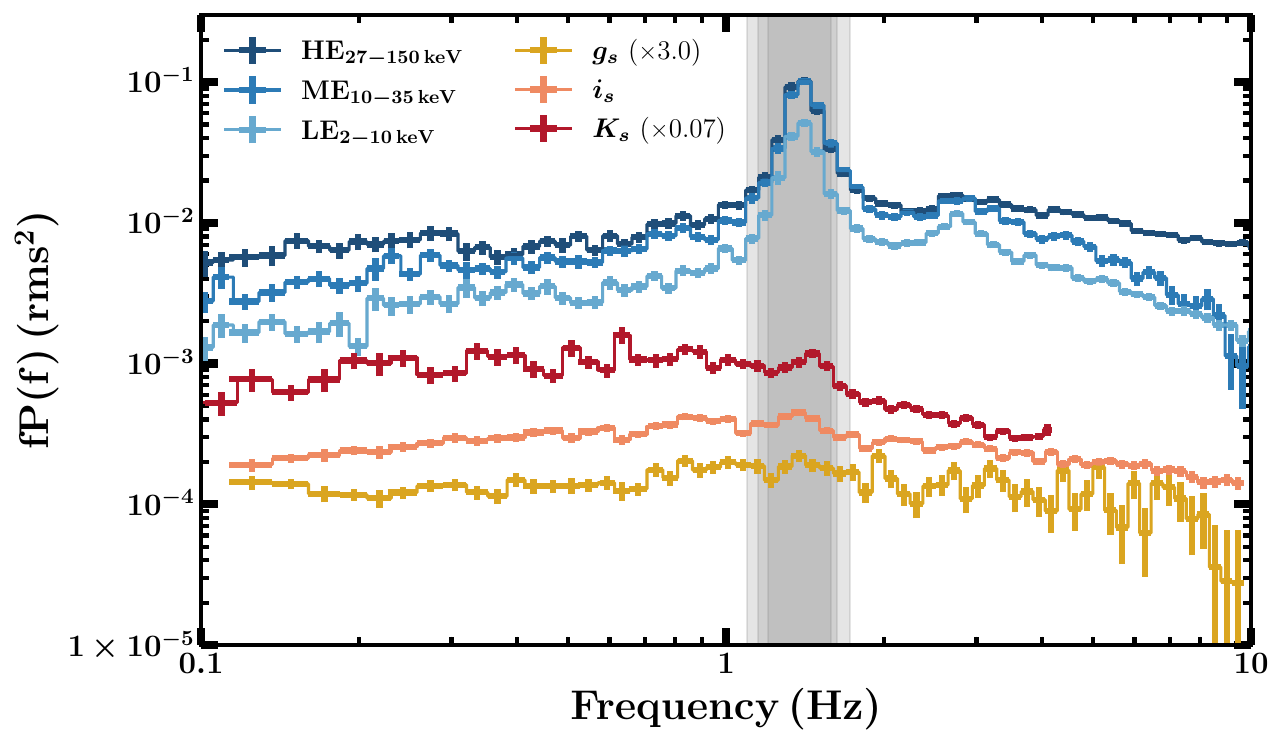} \\%column
\caption{PDS of Swift J1727.8--1613 in the X-ray (LE: 2--10 keV, ME: 10--35 keV, and HE: 27--150 keV; shown from light to dark blue), optical ($g_s$ and $i_s$; yellow and orange), and NIR ($K_s$; red) bands. For visual clarity, the $g_s$ and $K_s$ PDS are multiplied by factors of 3 and 0.07, respectively. The grey shaded region marks the QPO frequency range ($v_c \pm$FWHM) derived from the X-ray data.}
\label{fig:pds}
\end{figure}
%%%%%%%%%%%%%%%%%%%%%%%%%%%%%%%%%%%%%%%%%%%%%%%%%%%%%%%%%%%%%%%%%%%%%%%%%

\section{Data analysis and Results} 
\label{sec:results}

\subsection{Power density spectrum}

We computed PDS in the optical ($i_s$ and $g_s$), NIR ($K_s$) and X-ray (LE: 2--10\,keV, ME: 10--35\,keV and HE: 27--150\,keV) bands. The $i_s$ and $g_s$ data have time resolutions of 16\,ms, corresponding to a Nyquist frequency of 31.25\,Hz. Using 64\,s FFT segments, the lowest sampled frequency for the optical data is 0.015625\,Hz. 
For the $K_s$ data, we adopted a time resolution of 125\,ms and 64\,s FFT segments, giving a Nyquist frequency of 4\,Hz and a lowest sampled frequency of 0.015625\,Hz.
For the X-ray data, the corresponding Nyquist and lowest frequencies are 16\,Hz and 0.0625\,Hz, respectively. Poisson noise was subtracted from all PDS\footnote{The Poisson noise level was subtracted using XRONOS \textit{powspec} (norm = $-2$) for the X-ray data. For the IR data, the noise level was estimated directly from the mean-squared photometric uncertainties of the light curve. For the optical data, the noise level was estimated from the high-frequency power of the rebinned power spectrum.}, which were then normalized to rms$^2$ Hz$^{-1}$ \citep{Belloni1990, Miyamoto1992}.

As shown in Fig.~\ref{fig:pds}, the $g_s$ and $K_s$ PDS are vertically shifted by factors of 3.0 and 0.07, respectively, for visual clarity only. 
%We fitted the PDS in each band with multiple Lorentzian components. 
The PDS in each band was modeled using multiple Lorentzian components, following the standard multi-Lorentzian decomposition for BHXBs \citep[e.g.,][]{Belloni2002}. The QPO significance was quantified as the ratio of the best-fit Lorentzian normalization to its 1$\sigma$ uncertainty (norm/$\sigma_{\rm norm}$).
A strong QPO feature is clearly detected in all X-ray bands, with significances of $13.7\sigma$ (LE), $13.4\sigma$ (ME), and $17.2\sigma$ (HE). In the OIR bands, the QPO is marginally detected in the $K_s$ and $i_s$ bands with significances of $3.0\sigma$ and $3.3\sigma$, while only a weak excess is present in the $g_s$, with significance of $1.2\sigma$.
The QPO centroid frequencies are measured to be $1.36\pm0.04$\,Hz in the $i_s$ band and $1.392\pm0.008$\,Hz in the LE band, consistent within the measurement uncertainties. The QPO frequency range is highlighted in Fig.~\ref{fig:pds} by grey-shaded regions corresponding to $\nu_c \pm \mathrm{FWHM}$, where FWHM denotes the full width at half maximum.

%%%%%%%%%%%%%%%%%%%%%%%%%%%%%%%%%%%%%%%%%%%%%%%%%%%%%%%%%%%%%%%%%%%%%%%%%
\begin{figure}
\centering
\includegraphics[width=\columnwidth]{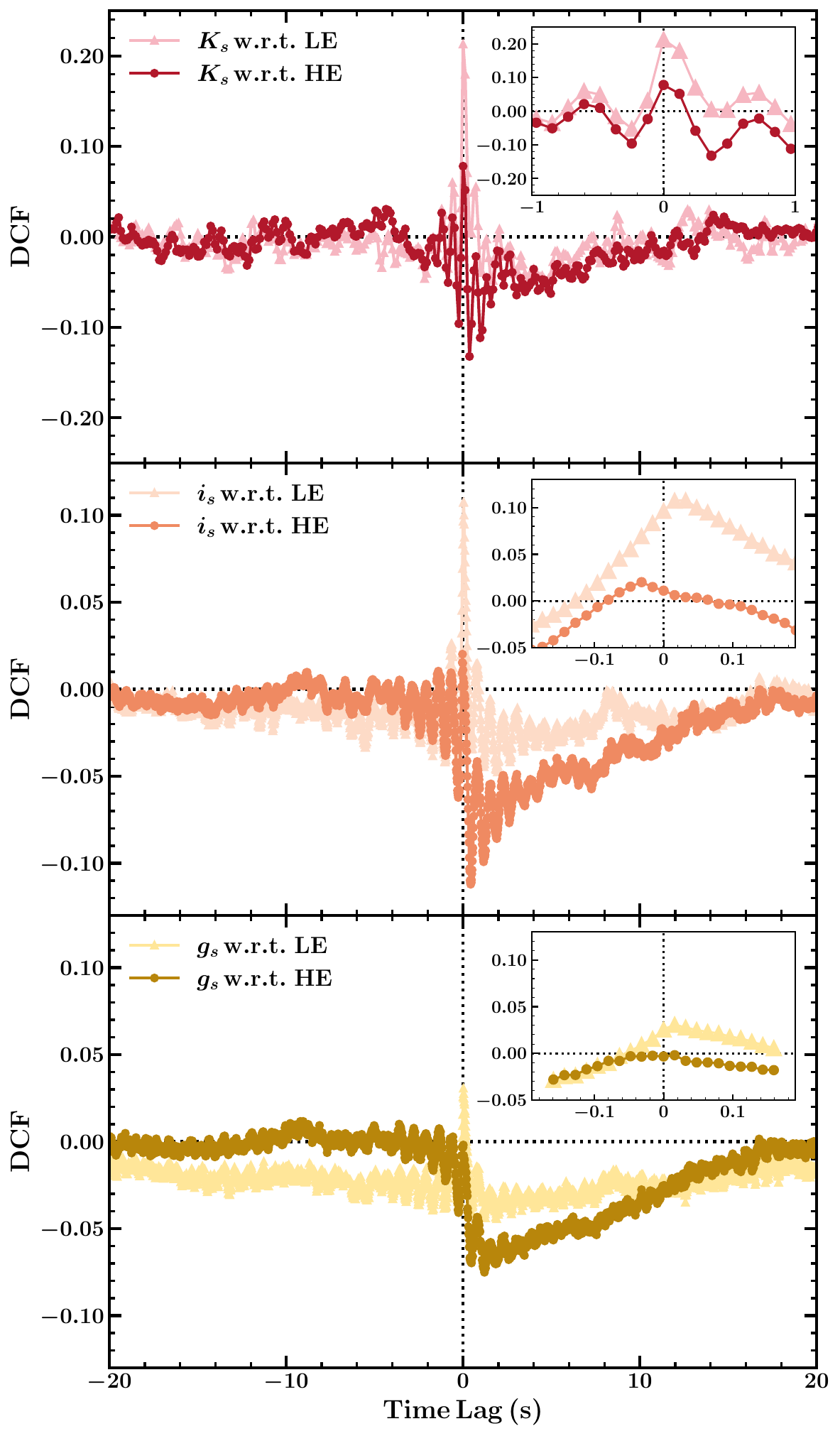} \\%column
\caption{From top to bottom, the DCFs of the $K_s$, $i_s$, and $g_s$ bands \textit{w.r.t.} the LE (2--10\,keV) and HE (27--150\,keV) bands of Swift J1727.8--1613 are shown. The correlations with the LE and HE bands are marked by light/dark red triangles/dots, light/dark orange triangles/dots, and light/dark yellow triangles/dots, respectively. In all panels, the X-ray band is taken as the reference band. The inset panels show zoomed-in views of the central peaks. A strong anti-correlation is observed in the $i_s$ and $g_s$ bands when the HE band is used as the reference band. }
\label{fig:ccf_oir}
\end{figure}
%%%%%%%%%%%%%%%%%%%%%%%%%%%%%%%%%%%%%%%%%%%%%%%%%%%%%%%%%%%%%%%%%%%%%%%%%

\subsection{Discrete correlation function}

To quantify the time lag and correlation between the OIR and X-ray light curves, we employed the discrete correlation function (DCF) following the method described in \citet{Edelson1988} and \citet{Gandhi2010}. 
Before calculating the DCF, all OIR and X-ray light curves were expressed in MJD after barycentric correction using the same source coordinates (see Section~\ref{sec:data_red} for more details). For the ULTRACAM and HAWK-I data, the time assigned to each photometric point corresponds to the mid-exposure time. For HXMT, light curves in different energy bands were extracted from the barycentre-corrected event files, ensuring a consistent time reference among the X-ray bands. We then calculated the DCF using the original time series, without interpolating one band onto the other. For each pair of data points from two bands, the time delay was defined as $\tau = t_{\rm OIR} - t_{\rm X}$, where the two times are the barycentric MJD timestamps of the paired data points. A positive lag therefore indicates that the OIR photons lag behind the X-rays.
In our analysis, the LE and HE bands were used as the reference bands, and the OIR data as the interest bands. 
The DCF values were normalized by the data standard deviations, ensuring that they lie within the range $[-1,1]$. The DCF was calculated from the original light curves, with time resolutions of 16\,ms for the optical band and 125\,ms for the NIR band. 
The short-timescale lag was estimated from the centre of the bin with the maximum DCF value, and is therefore subject to a binning uncertainty of approximately half a bin.

As shown in Fig.~\ref{fig:ccf_oir}, we present the DCFs of the $K_s$ (red), $i_s$ (orange) and $g_s$ (yellow) bands with respect to (\textit{w.r.t}) the LE (2--10\,keV) and HE (27--150\,keV) bands. 
Using the LE band as the reference, all three OIR bands show a strong positive correlation with the X-ray emission, consistent with the results reported by \citet{Vincentelli2025}. However, when the reference band is shifted to the harder HE band, the correlation behaviour changes significantly. While the $K_s$ band still shows a strong positive correlation with the HE data, the $i_s$ and $g_s$ bands instead exhibit a strong anti-correlation, with the optical emission lagging behind the hard X-rays.
In addition, the DCFs show an oscillatory structure around zero lag, with a characteristic period of $\sim0.7-0.8$\,s. This is broadly consistent with the QPO period inferred from the PDS, $P_{\rm QPO}=1/\nu_{\rm QPO}\sim 0.72-0.74$\,s, suggesting that the DCF oscillations are likely associated with the QPO modulation. 

The inset panels of Fig.~\ref{fig:ccf_oir} show the DCFs on short timescales around zero lag. No significant short-timescale lag is detected in the $K_s$ band, likely due to its relatively large time resolution of $\sim125$\,ms. In contrast, the $i_s$ photons lag the LE photons by $\sim32$\,ms, while leading the HE photons by $\sim32$\,ms. The $g_s$ photons lag the LE photons by $\sim16$\,ms, whereas no significant lag is detected between the $g_s$ and HE emission.

% %%%%%%%%%%%%%%%%%%%%%%%%%%%%%%%%%%%%%%%%%%%%%%%%%%%%%%%%%%%%%%%%%%%%%%%%%
\begin{figure}
\centering
\includegraphics[width=\columnwidth]{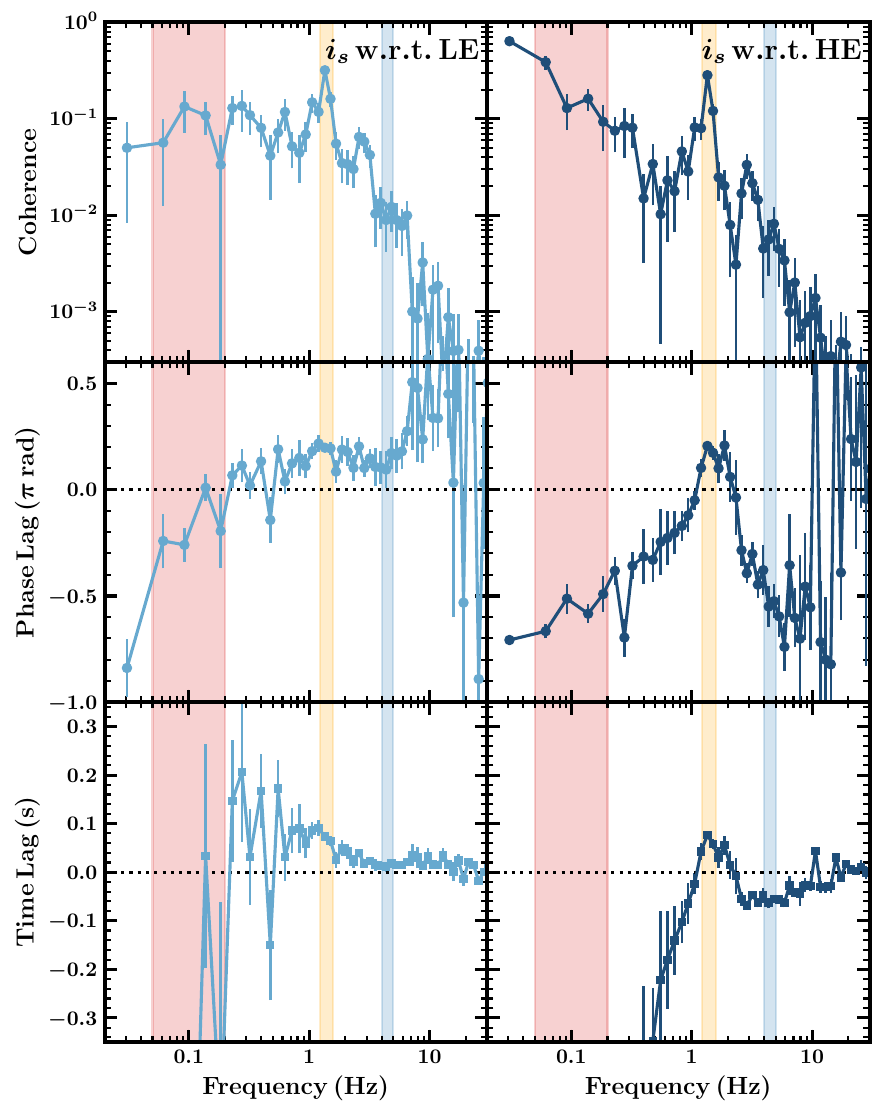} \\%column
\caption{Coherence (top panel), phase lag (middle panel) and time lag (bottom panel) of optical ($i_s$) \textit{w.r.t.} LE (2--10\,keV; left panel) and HE (27--150\,keV; right panel) bands for Swift J1727.8--1613. A positive lag denotes optical lagging X-rays. The orange, red, and blue shaded regions indicate the QPO frequency range ($\nu_c \pm $FWHM), the LFBN band (0.05--0.2\,Hz) and the HFBN band (4--5\,Hz), respectively. }
\label{fig:lag_o}
\end{figure}
%%%%%%%%%%%%%%%%%%%%%%%%%%%%%%%%%%%%%%%%%%%%%%%%%%%%%%%%%%%%%%%%%%%%%%%%%

\subsection{Cross spectra}

Since cross-spectral analysis enables us to measure the time lag as a function of Fourier frequency, it also allows us to distinguish variability with different characteristic timescales. We used {\sc stingray}\footnote{\url{https://docs.stingray.science/en/v2.0.0/index.html}} to calculate the cross spectra, adopting the HXMT bands as the reference. The $i_s$ and $g_s$ data were analyzed with a time resolution of 16\,ms and divided into 32.768\,s segments, while the $K_s$ data were analyzed with a time resolution of 125\,ms and 32\,s segments. 
The cross spectra were logarithmically rebinned with a factor of 0.1 and normalized using the rms normalization \citep{Belloni1990,Miyamoto1992}. 
For the main cross-spectral analysis, we focus on the $i_s$ band, which provides the highest-quality OIR timing data. The $g_s$ band shows similar behaviour, while the $K_s$ results (Fig.~\ref{fig:lag_ir}) are limited by lower time resolution and shorter overlap with HXMT, as also applies to Section~\ref{sec:lag_ene}.

In Fig.~\ref{fig:lag_o}, we present the cross-spectra of the $i_s$ band \textit{w.r.t} the LE and HE bands. The orange, red, and blue shaded regions indicate the QPO, low-frequency broadband noise (LFBN; 0.05--0.2\,Hz), and high-frequency broadband noise (HFBN; 4--5\,Hz), respectively. The LFBN and HFBN frequency intervals are selected to minimise contamination from the QPO fundamental and harmonic components, while ensuring reliable lag measurements at low frequencies and sufficient coherence at high frequencies (see the HE phase-lag and coherence spectra in Fig.~\ref{fig:lag_o}). A significant coherence is detected at the QPO centroid frequency, consistent with the results reported by \citet{Vincentelli2025}.

We compute the QPO time (phase) lag over the frequency range $\nu_c \pm \mathrm{FWHM}$ (1.22--1.58\,Hz). The $i_s$ band is found to lag the LE band by $71\pm7$\,ms ($0.19\pm0.02\,\pi$\,rad), and the HE band by $67\pm6$\,ms ($0.18\pm0.02\,\pi$\,rad), respectively.
Furthermore, when using LE as the reference band, the $i_s$ photons lead the LE photons in the LFBN component, while lagging behind them in the HFBN component. In contrast, when using HE as the reference band, the $i_s$ photons consistently lead the HE photons in both the LFBN and HFBN components. 
The lag spectra obtained using HE as the reference band may also be interpreted as consisting of a QPO phase lag superimposed on an underlying broadband lag component.
A more detailed investigation of the lag--energy dependence is presented in Section~\ref{sec:lag_ene}.

%%%%%%%%%%%%%%%%%%%%%%%%%%%%%%%%%%%%%%%%%%%%%%%%%%%%%%%%%%%%%%%%%%%%%%%%%
\begin{figure*}
\centering
\includegraphics[width=\textwidth]{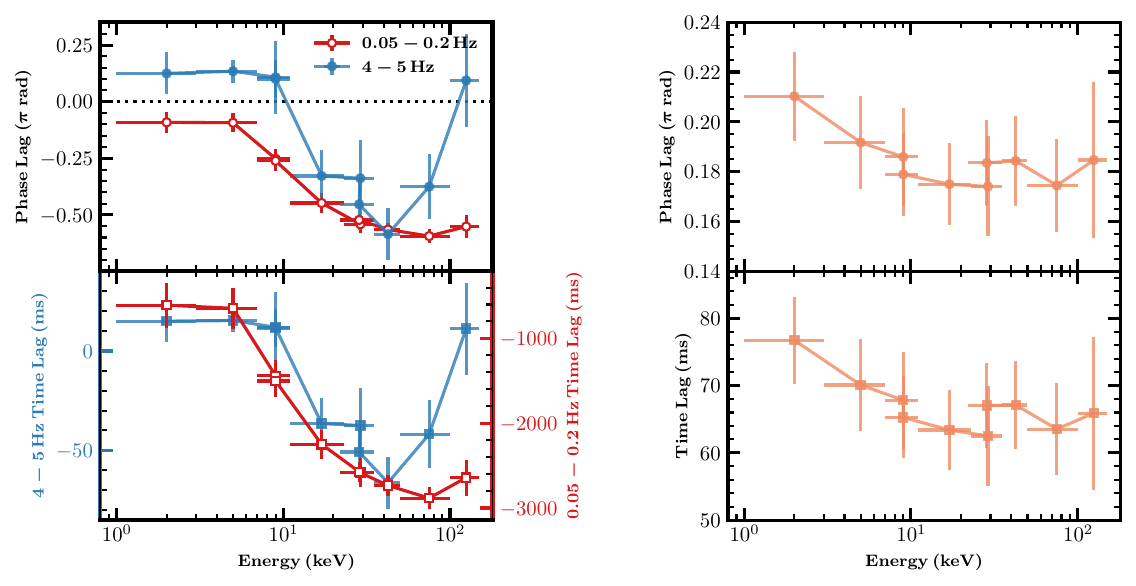} \\%column
\caption{Phase-lag (top panel) and time-lag (bottom panel) spectra of the optical ($i_s$) band \textit{w.r.t} the X-ray band (2--150\,keV) for Swift J1727.8--1613. The left panels show the results for the LFBN (red open symbols; 0.05--0.2\,Hz) and HFBN (blue filled symbols; 4--5\,Hz), while the right panels present the corresponding QPO (1.22--1.58\,Hz) results. 
A positive lag denotes optical lagging X-rays. }
\label{fig:lagpha_bnqpo_o}
\end{figure*}
%%%%%%%%%%%%%%%%%%%%%%%%%%%%%%%%%%%%%%%%%%%%%%%%%%%%%%%%%%%%%%%%%%%%%%%%%

\subsection{Lag-energy spectra}
\label{sec:lag_ene}

To investigate the energy dependence of lags, we divided the X-ray bands into several sub-bands: LE (1--3, 3--7, 7--11\,keV), ME (7--11, 11--23, 23--35\,keV), and HE (22--35, 35--50, 50--100, 100--150\,keV)\footnote{To ensure consistency between the instruments, we adopted overlapping energy bands between LE and ME, as well as between ME and HE.}, and computed the lag-energy spectra. We took the X-ray data as the reference band; therefore, a positive lag indicates that the optical emission lags behind the X-rays.

In the left panel of Fig.~\ref{fig:lagpha_bnqpo_o}, we present the lag-energy spectra between the $i_s$ band and the X-ray data for the LFBN (0.05--0.2 Hz) and HFBN (4--5 Hz) frequency ranges.
For the LFBN, the optical photons consistently lead the X-ray photons. The phase (time) lag remains approximately $-0.1\pi$\,rad ($\sim$0.5\,s) within the 1--7\,keV range, and gradually shifts to $\sim-0.6\pi$\,rad ($\sim$3\,s) as the X-ray energy increases up to 100\,keV. Above this energy, the optical lag shows a slight decrease to $\sim -0.55\pi$\,rad ($\sim$2.6\,s).
For the HFBN, the energy dependence of the lag is more complex, as shown in the left panel of Fig.~\ref{fig:lagpha_bnqpo_o}. Below 10\,keV, the optical photons lag behind the X-ray photons, with a positive lag of $\sim0.12\pi$\,rad ($\sim15$\,ms). The lag then decreases with energy, crossing zero and turning into a lead that reaches $\sim-0.6\pi$\,rad ($\sim75$\,ms) at $\sim50$\,keV. At higher energies, the lead weakens to $\sim-0.4\pi$\,rad ($\sim40$\,ms) before further dropping to $\sim-0.1\pi$\,rad ($\sim11$\,ms) above 100\,keV.

We also computed the lag-energy spectrum in the QPO frequency range (1.22--1.58\,Hz) between the $i_s$ and X-ray bands, as shown in the right panel of Fig.~\ref{fig:lagpha_bnqpo_o}. The $i_s$ photons consistently lag behind the X-ray photons across the full energy range. As the X-ray energy increases from 2\,keV to 23\,keV, the optical lag slightly decreases from $0.21\pi$\,rad ($\sim77$\,ms) to $0.175\pi$\,rad ($\sim63$\,ms). At higher X-ray energies, up to 150\,keV, the optical lag shows a slight increase again, reaching $\sim0.185\pi$\,rad ($\sim66$\,ms).

\section{Discussion}

We performed a simultaneous OIR ($g_s$, $i_s$, and $K_s$) and broadband X-ray (2--150 keV) timing study of Swift J1727.8--1613 in the hard-intermediate state. The CCFs reveal a strong energy- and wavelength-dependent coupling: all three OIR bands are positively correlated with the soft X-rays, whereas the optical bands ($g_s$ and $i_s$) show a strong anti-correlation with the hard X-rays, reported here for the first time. On short timescales, the $i_s$ photons lag the LE photons by $\sim32$ ms but lead the HE photons by a similar amount, while the $K_s$ band remains positively correlated with the hard X-rays.
The frequency-resolved analysis reveals distinct energy-dependent lag behaviours between the broadband-noise and QPO components. The LFBN is characterized by a persistent $i_s$ lead that becomes stronger towards higher X-ray energies. The HFBN shows a stronger energy dependence: $i_s$ lags the soft X-rays but leads the hard X-rays, with the hard-band lead weakening again at the highest energies. In contrast, the QPO lag--energy spectrum is nearly flat, with an approximately constant $i_s$ lag of $\sim60$--80 ms ($\sim0.18\pi$ rad) across the full 2--150 keV band.

In the following subsections, we discuss the origin of these complex OIR/X-ray correlations, with particular emphasis on the newly identified hard X-ray anti-correlation, and then examine the physical origin of the simultaneous X-ray and optical QPOs.

\subsection{Origin of the complex cross-correlation}

It is not the first time that simultaneous sub-second X-ray and OIR observations reveal an anti-correlation between these two bands in BHLMXBs, such as Swift J1753.5--0127 \citep{Durant2008,Hynes2009,Veledina2017}, MAXI J1535--571  \citep{Vincentelli2021} and MAXI J1820+070 \citep[][]{Paice2019, Paice2021}. The jet internal shock model could explain such an anti-correlation through variability induced by fluctuations in Doppler-boosted jet emission \citep{Malzac2018}. In this scenario, fast moving shells collide with preceding slower ejecta, producing internal shocks within the jet, which then  naturally gives rise to delayed IR anti-correlated variability and sub-second OIR lags \citep[e.g.,][]{Casella2010, Malzac2014, Malzac2018}.
However, the jet internal-shock scenario is not expected to exhibit an energy dependence, as it should be independent of the spectral component.
The qualitatively similar, albeit weaker, anti-correlation trend observed in the LE DCF suggests that an anti-correlated component may also contribute at softer X-ray energies, although it is likely diluted by the disk emission. This behaviour may point to a hybrid scenario in which jet-related variability is coupled to a hot flow.

\begin{figure}
\centering
\includegraphics[width=1\columnwidth]{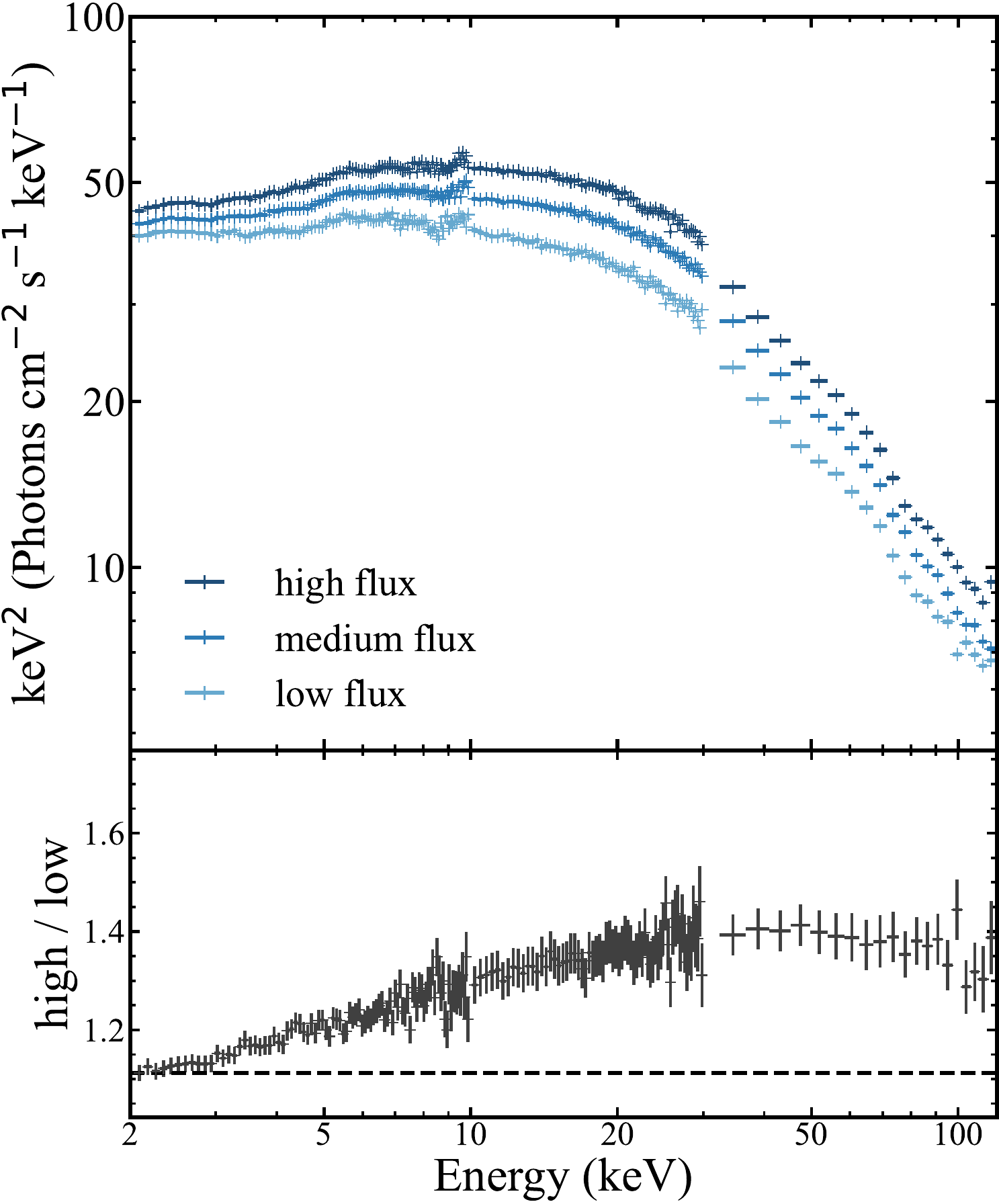} \\%column
\caption{Top panel: Model-unfolded spectra of Swift J1727.8--1613 observed with HXMT in the 2--150 keV band at different flux levels. The data were divided into three flux intervals based on the median count rate. The light, medium, and dark blue spectra correspond to the low-, medium-, and high-flux states, respectively. Bottom panel: Spectral ratio between the high- and low-flux states.}
\label{fig:pha_hxmt}
\end{figure}

A hot flow origin offers an alternative explanation \citep{Veledina2013a}. In this scenario, broadband emission is produced by the synchrotron self-Compton (SSC) mechanism from hybrid electrons. 
The optical part of the spectrum is dominated by synchrotron emission of the non-thermal component of the electron distribution \citep[which dominates over the synchrotron emission of the thermal particles,][]{Wardzinski2001}. In contrast, the X-ray continuum is dominated by thermal Comptonisation by the Maxwellian electrons, while the direct Comptonisation contribution from the non-thermal electrons is expected to be minor \citep[e.g.,][]{Poutanen2009, Veledina2013a}.
Anti-correlation arises from increased synchrotron self-absorption at higher X-ray luminosity, causing the overall spectrum to pivot \citep{Veledina2011}.
Our analysis of spectral variations on second timescales, based on 1-s time bins divided into three flux intervals, reveals a ``harder-when-brighter'' behaviour: the high-to-low flux spectral ratio increases with energy (see Fig.~\ref{fig:pha_hxmt}), and the hardness ratio (30--100 keV/2--5 keV) shows a positive correlation with intensity, increasing from 0.68 to 0.77 and 0.82 across the three flux levels.

The anti-correlation dip shifted to positive optical lags has previously been observed in the Swift J1753.5--0127 and MAXI J1535--571 data \citep{Hynes2009,Veledina2017,Vincentelli2021} and its explanation involved additional contribution from disk Comptonisation (DC) component.
Interestingly, broadband spectral studies of Swift J1727.8--1613 suggest that the hot flow cannot be described by a single Comptonisation component alone, but instead requires at least two Comptonisation regions \citep[e.g.,][]{Liu2024, Ma2025,Chand2026}. 
Moreover, a scenario involving multiple Comptonisation components has also been proposed for other BHXBs. For example, in MAXI J1820+070, \citet{Yang2026} used the observed incoherence at high X-ray energies to propose a two-Comptonisation framework that accounts for both the short- and long-timescale variability.
In this context, the stronger anti-correlation observed in the hard X-ray band may indicate that the anti-correlated variability is more directly associated with the harder of these Comptonized components, potentially linked to SSC emission from the hot flow. In contrast, the weaker anti-correlation in the LE band may reflect dilution by softer emission components, such as the disk and/or a separate, softer Comptonisation region.

If such an anti-correlated component is present across a broad range of Fourier frequencies, it could naturally produce an approximately constant phase lag close to $-0.5\pi$\,rad (see right panel of Fig.~\ref{fig:lag_o} and also \citealt{Vincentelli2021}). In this picture, the deviation from the nearly constant phase lag around the QPO frequency ($\approx1$\,Hz) could reflect the superposition of an additional coherent QPO modulation, arising from a different variability process (see next section), on top of the underlying broadband anti-correlated component. 
This interpretation is also qualitatively consistent with recent broadband X-ray spectral-timing studies on the same source, which suggest that QPO and broadband-noise variability can contribute differently to the measured cross-spectral lags even at the same Fourier frequencies \citep{Bollemeijer2025}. 
Interestingly, timing analyses revealed a pronounced excess in the QPO rms spectrum above $\sim20$--$40$\,keV \citep{Yang2024}, while \citet{Jin2025} also reported ``a high-energy excess'' in the high-energy PDS of the same source.
In this framework, the short negative optical lag observed in the hard X-ray DCF may arise from the combined contribution of the broadband anti-correlated variability and the additional QPO modulation.

\begin{figure}
\centering
\includegraphics[width=0.9\columnwidth]{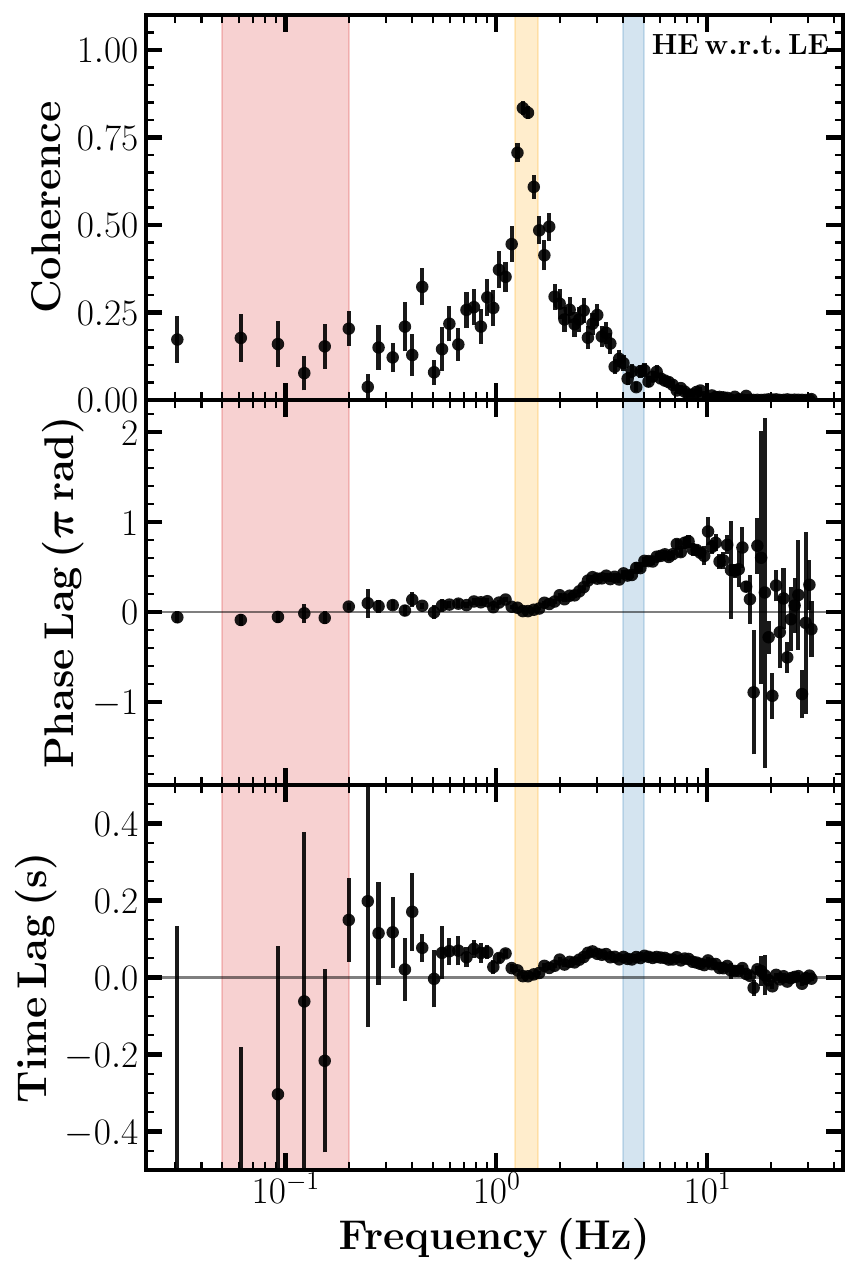} \\%column
\caption{Similar to Fig.~\ref{fig:lag_o}, but showing the cross-spectra of the HE (27--150\,keV) \textit{w.r.t.} LE (2--10\,keV) data in Swift J1727.8--1613.
A positive lag denotes hard X-rays lagging soft X-rays.}
\label{fig:lag_hxmt}
\end{figure}

We also note that the X-ray cross-spectral analysis shows that the hard X-ray photons lag the soft X-ray photons (Fig.~\ref{fig:lag_hxmt}). 
For the two-component Comptonisation, these are associated with the phase lags, rather than with the physical time delays, and can be as high as $\Delta\phi\sim\pi$, i.e. $\Delta t\sim1/(2f)$ \citep[where $f$ is the Fourier frequency;][]{Veledina2018}.
The dependence of the phase lags on energy is likewise well in line with the expectations of the two-component Comptonisation \citep{Veledina2018}.

The dependence of the DCF on OIR wavelength is noteworthy.
The $K_s$/HE DCF clearly shows a sharp peak at a small delay, the feature that is typically associated with the jet \citep{Casella2010,Malzac2014,Gandhi2010,Gandhi2017,Paice2021}.
At the same time, the $g_s$ band does not seem to have this peak.
This might indicate the transition between the jet-dominated emission in the NIR towards a hot flow-dominated regime in the optical.

%%%%%%%%%%%%%%%%%%%%%%%%%%%%%%%%%%%%%%%%%%%%%%%%%%%%%%%%%%%%%%%%%%%%%%%%%
%%%%%%%%%%%%%%%%%%%%%%%%%%%%%%%%%%%%%%%%%%%%%%%%%%%%%%%%%%%%%%%%%%%%%%%%%

%%%%%%%%%%%%%%%%%%%%%%%%%%%%%%%%%%%%%%%%%%%%%%%%%%%%%%%%%%%%%%%%%%%%%%%%%
%%%%%%%%%%%%%%%%%%%%%%%%%%%%%%%%%%%%%%%%%%%%%%%%%%%%%%%%%%%%%%%%%%%%%%%%%

%%%%%%%%%%%%%%%%%%%%%%%%%%%%%%%%%%%%%%%%%%%%%%%%%%%%%%%%%%%%%%%%%%%%%%%%%
\begin{figure*}
\centering
\includegraphics[width=\textwidth]{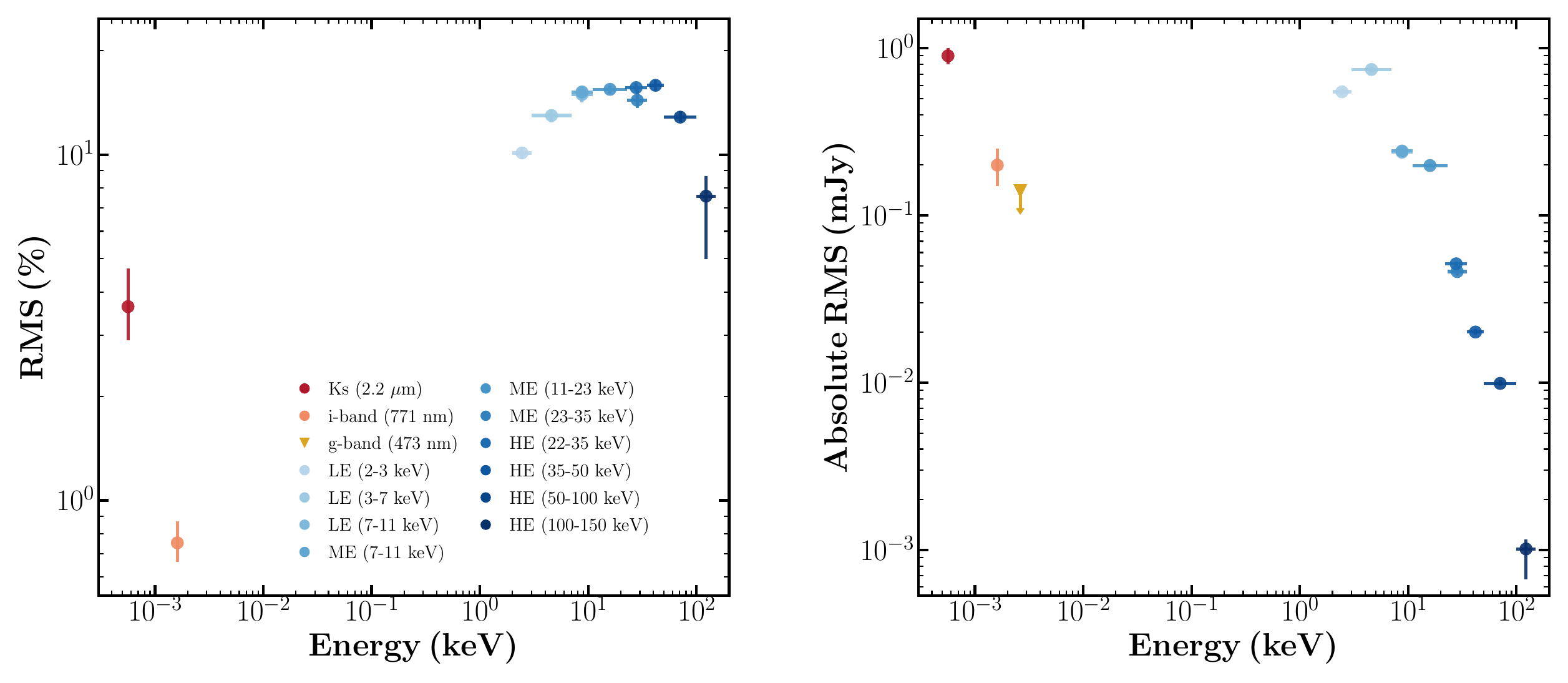} \\%column
\caption{QPO rms (left) and absolute rms (right) of Swift J1727.8--1613 in NIR ($K_s$), optical ($i_s$ and $g_s$), and X-ray (2--150\,keV). Since no significant QPO was detected in the $g_s$ band, only an upper limit on the QPO absolute rms is shown for this band.}
\label{fig:rms_all}
\end{figure*}
%%%%%%%%%%%%%%%%%%%%%%%%%%%%%%%%%%%%%%%%%%%%%%%%%%%%%%%%%%%%%%%%%%%%%%%%%

\subsection{X-ray and optical QPOs}

One of the most notable timing features of Swift J1727.8--1613 is the detection of QPOs in both the X-ray and OIR bands \citep[see also][]{Vincentelli2025}. This provides a valuable opportunity to investigate the physical origin of multi-wavelength QPOs and their connection to the inner accretion flow and/or jet activity.

To date, the hot-flow Lense–Thirring precession model has successfully explained the origin of X-ray QPOs in several studies of Swift J1727.8--1613. For example, \citet{Xu2025} showed that QPOs below and above 3 Hz exhibit different relationships with the disk, Comptonisation, and reflection components, and proposed that these behaviours can be naturally explained within the hot-flow Lense–Thirring precession framework.
\citet{Ma2025} found a tight correlation predicted by the model between the inner disk radius and QPO frequency, regardless of changes in the accretion rate, particularly during the flare state of Swift J1727.8--1613. 
In addition, the X-ray energy dependence also provides an important test of the precessing hot-flow scenario. Simulations by \citet{You2018} predict an increasing QPO rms with photon energy, as higher-energy photons undergo more Compton scatterings before escaping the hot flow. 
In contrast, we observe a decrease in QPO rms above $\sim$50 keV in this work (left panel of Fig.~\ref{fig:rms_all}). This discrepancy may indicate that the rms–energy relation is more complex and depends on the spectral state and the properties of the Comptonising flow. Previous observations have shown that the QPO rms–energy relation can remain approximately constant or increase at energies below $\sim$15 keV before flattening \citep[e.g.,][]{Yu2024, Xu2025}.

The hot-flow Lense–Thirring precession model attributes optical QPOs to emission from the outer regions of the precessing hot flow \citep{Veledina2013b}. 
In this framework, the fundamental QPO is expected to have a larger fractional rms amplitude in the optical band than in X-rays, since the optical emission originates from synchrotron radiation in the hot flow, which is more strongly dependent on the viewing angle.
Conversely, the harmonic component is predicted to be stronger in X-rays than in the optical band, owing to the more complex angular radiation pattern and relativistic effects, which can lead to a double-peaked X-ray light curve. 
In Swift J1727.8--1613, the detection of an X-ray harmonic QPO ($\sim3.6\sigma$ in the LE band), but not in the OIR bands, is not inconsistent with this prediction. However, it should be noted that the current OIR constraints are insufficient for a stringent test. For example, the $3\sigma$ upper limit in the $i_s$ band ($<0.38$\%) remains above the expected harmonic QPO rms ($\sim0.25$\%), estimated from the X-ray harmonic-to-fundamental rms ratio.
In addition, the X-ray QPO fundamental rms amplitude ($\sim8$--15\%) is substantially higher than that in the OIR bands ($<4$\%; see Fig.~\ref{fig:rms_all}), contrary to the predicted trend.
Hence, a combination of the variable hot flow component with a more stable, at the considered timescales, disk emission is needed.

An alternative interpretation is provided by the small-scale jet model, which proposes that the observed X-ray QPOs originate from the Lense–Thirring precession of a compact jet structure \citep{Malzac2018,Ma2021}. This framework also predicts the presence of simultaneous optical QPOs, offering a consistent explanation for both the X-ray QPOs observed up to 200 keV and their optical counterparts \citep{Ma2021, Thomas2022}. Within this model, the upper region of the jet is assumed to be less curved, resulting in a smaller optical QPO fractional rms amplitude.
Furthermore, the model predicts an optical phase delay of approximately $\pi$ between the X-ray–emitting and optical–emitting regions in MAXI J1820+070 \citep{Ma2021}. 
In comparison to MAXI J1820+070, Swift J1727.8--1613 shows a phase lag of $\sim0.19\pi$\,rad, corresponding to a sub-second optical lag of approximately 60--80 ms. This shorter optical lag may suggest that the source hosts a compact jet region, with a characteristic scale of $1.8\times10^3\,{\rm R_g}$ (where ${\rm R_g}=GM/c^2$).

In addition, a simple power-law fit to the X-ray QPO absolute-rms spectrum gives ${\rm AbsRMS}\propto E^{-1.61\pm0.14}$, suggesting a steep high-energy shape for the QPO-modulated component. 
Although this absolute-rms-energy slope is not directly equivalent to a time-averaged spectral index, it is much steeper than the optically thin jet-synchrotron slope inferred for XTE J1118+480 \citep[$F_{\nu}\propto\nu^{-0.8}$;][]{Markoff2001}. 
This disfavors a simple unbroken jet-like synchrotron power law as the dominant QPO-modulated hard X-ray component, and instead favours a steepened or cutoff Comptonised component \citep[e.g.][]{Zdziarski2004,Done2007}.
A jet-related contribution, however, cannot be excluded if it is coupled to the Comptonising region or affected by cooling/cutoff effects \citep{Russell2013}.

\section*{Acknowledgements}
We thank the anonymous referee for useful comments that helped us improve the paper. 
RM thanks C. Done, N. Castro Segura and Z. Xu for helpful discussions.
RM acknowledges support from the Royal Society Newton Funds. PG thanks support by the Science and Technology Facilities Council. 
FV and TS acknowledge financial support from the Spanish Ministry of Science, Innovation and Universities (MICIU) under grant PID2023-151588NB-I00.
AV acknowledges support from the Research Council of Finland grants 355672 and 372881. 
Nordita is supported in part by NordForsk. 
% The Acknowledgements section is not numbered. Here you can thank helpful
% colleagues, acknowledge funding agencies, telescopes and facilities used etc.

%%%%%%%%%%%%%%%%%%%%%%%%%%%%%%%%%%%%%%%%%%%%%%%%%%
\section*{Data Availability}

This work has made use of the data from the \textit{Insight}-HXMT mission, a project funded by China National Space Administration (CNSA) and the Chinese Academy of Sciences (CAS). All data are public and can be found at \url{http://archive.hxmt.cn/proposal.}
{\sc ultracam} data can be made available upon reasonable request to the authors. 
HAWK-I raw data are available on the ESO public archive.
 
% The inclusion of a Data Availability Statement is a requirement for articles published in MNRAS. Data Availability Statements provide a standardised format for readers to understand the availability of data underlying the research results described in the article. The statement may refer to original data generated in the course of the study or to third-party data analysed in the article. The statement should describe and provide means of access, where possible, by linking to the data or providing the required accession numbers for the relevant databases or DOIs.



\bibliographystyle{mnras}
\bibliography{ref} % if your bibtex file is called example.bib

%\begin{thebibliography}{99}
%\end{thebibliography}

%%%%%%%%%%%%%%%%%%%%%%%%%%%%%%%%%%%%%%%%%%%%%%%%%%

%%%%%%%%%%%%%%%%% APPENDICES %%%%%%%%%%%%%%%%%%%%%

\appendix

\section{Supplementary Figures}

% %%%%%%%%%%%%%%%%%%%%%%%%%%%%%%%%%%%%%%%%%%%%%%%%%%%%%%%%%%%%%%%%%%%%%%%%%
\begin{figure}
\centering
\includegraphics[width=\columnwidth]{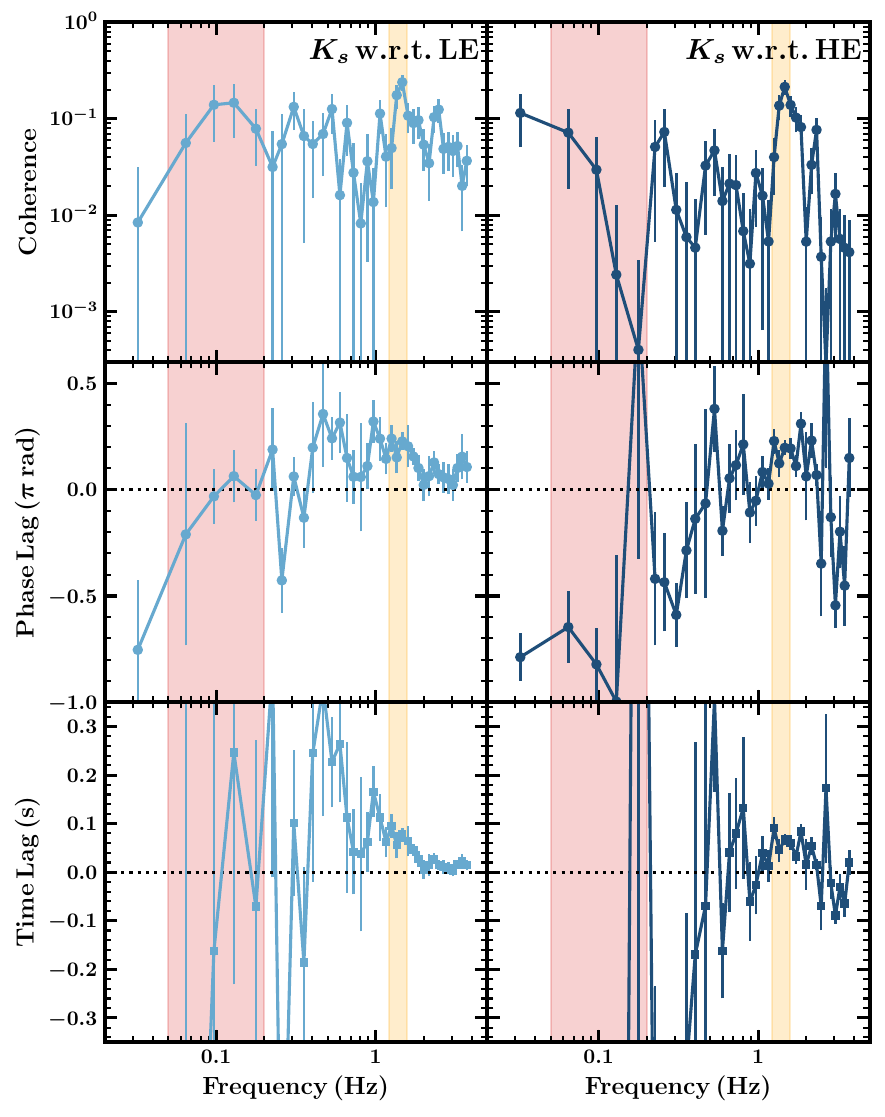} \\%column
\caption{Coherence (top panel), phase lag (middle panel) and time lag (bottom panel) of NIR ($K_s$) \textit{w.r.t.} LE (2--10\,keV; left panel) and HE (27--150\,keV; right panel) bands for Swift J1727.8--1613. A positive lag denotes NIR lagging X-rays. The orange and red shaded regions indicate the QPO frequency range ($\nu_c \pm $FWHM), and the LFBN band (0.05--0.2\,Hz), respectively. }
\label{fig:lag_ir}
\end{figure}
%%%%%%%%%%%%%%%%%%%%%%%%%%%%%%%%%%%%%%%%%%%%%%%%%%%%%%%%%%%%%%%%%%%%%%%%%

% Don't change these lines
\bsp	% typesetting comment
\label{lastpage}
\end{document}